\documentclass[12pt]{article}
\usepackage{a4,amsmath,lscape,epsfig}

\begin{document}

\begin{center}
{\bf Quons Restricted to the Antisymmetric Subspace: \\
Formalism and Applications}
\end{center}

\begin{center}
{\it S.S. Avancini, F.F. de Souza Cruz, J.R. Marinelli, \\
D.P.Menezes and M.M. Watanabe de Moraes}\\
{\it Depto de F\'{\i}sica - CFM - Universidade Federal de Santa
Catarina}\\
{\it Florian\'opolis - SC - CP. 476 - CEP 88.040 - 900 - Brazil}
\end{center}

\vspace{0.50cm}

\begin{abstract}
In this work we develop a formalism to treat quons restricted to
the antisymmetric part of their many-body space. A model in which
a system of identical quons interact through a pairing force is
then solved within this restriction  and the differences between
our solution and the usual fermionic model solution are then
presented and discussed in detail. Possible connections to
physical systems are also considered.
\end{abstract}

\vspace{0.50cm} PACS number(s):03.65.Fd;21.60.Fw

\section {Introduction}

Quons are particles bearing statistics which interpolate between
the fermionic and the bosonic ones, depending on a deformation
parameter \cite{quons} $q$ which is defined in the interval
between $-1$ (fermionic limit) and $+1$ (bosonic limit). It can be
shown \cite{quons,sidric} that the space spanned by a system of
$n$ identical quons can be separated into subspaces according to
their permutational properties, that is, the symmetric,
antisymmetric and mixed symmetry subspaces. The restriction of
the dynamics to the symmetric subspace  was recently investigated
\cite{sidric}. Applications were done for the harmonic oscillator
and the rotor hamiltonians  and the results compared to the better
known deformed algebra ones. Besides the intrinsic interest on
these applications, the projection of the dynamics onto the
symmetric subspace brings many formal simplifications and has a
great advantage as compared with the solution  of boson-like
systems in the whole quonic space. This is the case for a system
of bosons that are in fact composed by  fundamental  fermions and
can display deviations from a true bosonic behavior under certain
conditions. A good example are the excitons in the high density
regime \cite{keldysh},
whose deviations come from the fermionic character
of the underlying particles, which can influence the dynamics if
the system enters in a regime where the fermion
occupation scheme become important. In principle, we may describe
such systems in a natural way within the quon algebra, if we keep
the q-values close (but not equal) to $+1$.

In the present work we concentrate our attention in the other
limit of the interval, i.e., the region of $q$-values close to $-1$.
In other words we restrict the dynamics to the antisymmetric subspace,
useful when the particles that
form the system have half-integer spin and so are fermion-like
particles. We then keep the system close to the fermionic
behavior, while we expect to cope with possible deviations through
the deformation parameter. In section 2, we show the
fundamental steps of our formalism for the restriction to the antisymmetric
space, which follows the same lines as discussed in \cite{sidric} for the
symmetric case.

As an application, we then consider a  pairing interaction model
\cite{cambia}, in its one and two level versions. This pairing
model, albeit a simple one, exhibits many rich features to test
our formalism and has already been investigated with the help of
quantum and quon algebras \cite{nossos}. Our main results are then presented
in section 3,
together with a careful analysis of the energy gap behavior,
characteristic of the pairing interaction. Specific physical cases in which
the effects studied here
are of interest will be the subject of future investigations.

\section{Antisymmetric quon states}
In this section we  build the antisymmetric basis states for a
system of $n$ quons and obtain the action of the annihilation
operator of a quon on these states. This operation is
crucial to obtain matrix elements of many-body operators. As it is
well known the quon algebra is defined as
\begin{equation}
a_i a^{\dag}_j - q a^{\dag}_j a_i = \delta_{ij},
\label{qalgebra}
\end{equation}
while the bare vacuum  $|0>$ satisfies the  condition $a_i |0>=0$.
An important quantity in this context is the deformed number,
defined as
\begin{equation}
[n]_q= \frac{1-q^n}{1-q}.
\end{equation}

 As far as we restrict our analysis to the antisymmetric subspace
we will need the {\it anti-deformed quantum number}, hereafter defined as
 \begin{equation}
\{n \}_q= [n]_{-q}=\frac{1-(-q)^n}{1+q}.
\label{anticaixote}
\end{equation}
In order to simplify our notation in future expressions we
suppress the parameter $q$ from the above notation, considering it
implicitly. Another definition to be used ahead is,

\begin{equation}
\{n\}!=\{n\}\{n-1\}....\{0\}!,
\end{equation}
with $\{0\}!=1$.

Following \cite{sidric} we  build the antisymmetric state for $n$
quons through an inductive procedure, starting from the
antisymmetric state for $n=2$ and $n=3$. In the first case, the
normalized antisymmetric state can be written as
\begin{equation}
|i~j;A>_{\cal N}= \frac{1}{\sqrt{2(1-q)}}(a^{\dag}_i a^{\dag}_j -
a^{\dag}_j a^{\dag}_i) |0>.
\end{equation}

Note that the subscript  ${\cal N}$ means that the state is
already normalized and the $A$ inside the ket means that the state
is completely antisymmetric under permutation of any particle
label. The action of the annihilation operator on the above state
gives
\begin{equation}
a_i |i~j;A>_{\cal N}= \frac{1}{\sqrt{2\{2\}}} (a_i a^{\dag}_i
a^{\dag}_j - a_i a^{\dag}_j a^{\dag}_i) |0>
=\sqrt{\frac{\{2\}}{2}} a^{\dag}_j |0>,
\end{equation}
where equation (\ref{qalgebra}) and the fact that $i \ne j$ were used.

Following the same steps, we can show that for $n=3$,
$$
a_i |i~j~k;A>_{\cal N}=\frac{1}{\sqrt{3!\{3\}!}}  a_i (a^{\dag}_i
a^{\dag}_j a^{\dag}_k - a^{\dag}_i a^{\dag}_k a^{\dag}_j +
a^{\dag}_k a^{\dag}_i a^{\dag}_j $$
$$
- a^{\dag}_k a^{\dag}_j a^{\dag}_i
+ a^{\dag}_j a^{\dag}_k a^{\dag}_i
- a^{\dag}_j a^{\dag}_i a^{\dag}_k)|0> $$
or simply
\begin{equation}
a_i |i~j~k;A>_{\cal N}= \sqrt{\frac{ \{3\}}{3}} |j~k;A>_{\cal N}.
\end{equation}

Continuing for larger numbers of quons, we define
\begin {equation}
|i_1~i_2...i_n;A>={\cal
A}a^{\dag}_{i_1}a^{\dag}_{i_2}...a^{\dag}_{i_n}|0>,
\end {equation}
with ${\cal A}$ being the usual anti-symmetrizer
operator which makes all the permutations of quon operators with
the appropriate sign. We then finally find,
\begin{equation}
a_l |i_1~i_2~i_3~...i_n;A>_{\cal N}= \sqrt {\frac {\{n\}}{n}}
|i_1~i_2~i_3~..i_{l-1}~i_{l+1}....i_n;A>_{\cal N} . \label{def}
\end{equation}

The above equations have a complete analogy with the ones
obtained in \cite{sidric} for the symmetric quonic state and the
general proof \cite{Duzzioni} follows the same lines presented in
the Appendix of that reference. With this result in hand we may
now obtain any necessary matrix element in order to get the
observables of the theory.

\section{Applications to the Pairing Model}

    Next, we want to apply the above results to solve the pairing
model  in a completely antisymmetric quonic basis. We  analyze the
one  and two level pairing model. We present bellow  the two
level hamiltonian which can easily be reduced to the one level
case. The model that we consider \cite{cambia} consists of two
$\Omega$ fold degenerate levels ($\Omega = j+1/2$) whose energy
difference is $\epsilon $. All the particles have angular momentum
$j$, the lower level has single-particle states labelled $j m_2$
and the upper level has single-particle states labelled $j m_1$.
The pairing Hamiltonian then reads:
\begin{equation}
H=\epsilon (\hat N_1 - \hat N_2) - \frac{G \Omega}{2} \left[
(\hat A^{\dag}_1 + \hat A^{\dag}_2)(\hat A_1+ \hat A_2) + (\hat A_1+ \hat
A_2)
(\hat A^{\dag}_1 + \hat A^{\dag}_2)
\right],
\label{hpair1}
\end{equation}
where $N_k$ is half the usual number operator, and measures the
number of pairs in each level $\{k\}$. We then have,
\begin{equation}
\hat N_k= \frac{1}{2} \sum_{m_k} a^{\dag}_{m_k} a_{m_k},
\end{equation}
and $\hat A^{\dag}_k$ creates a zero angular momentum pair
\begin{equation}
\hat A^{\dag}_k =\frac{1}{2 \sqrt \Omega} \sum_{m_k} (-1)^{j-m_k}
a^{\dag}_{m_k} a^{\dag}_{-m_k}, \label{Azao}
\end{equation}

\noindent where $k=1,2$.

This model originally built for fermions is modified when the
quon commutation rule is imposed. This can be explicitly seem when
we rewrite the hamiltonian (\ref{hpair1}), using equation
(\ref{qalgebra}),

\begin{equation}
H=\epsilon (\hat N_1 - \hat N_2) - \frac{G \Omega}{2} \left[\hat
A^{\dag}_1
\hat A_1 + \hat A^{\dag}_2 \hat A_2 + \hat A^{\dag}_1 \hat A_2 + \hat
A^{\dag}_2 \hat A_1 \right](1+q^4) + H',
\label{hpair2}
\end{equation}
and where

\begin{equation}
H'= \frac{G}{8} (q-1)[4 \Omega + 2(q-q^2)(\hat N_1 + \hat N_2)].
\label{hlinha}
\end{equation}

Notice that the modifications come from  the commutation rules
and reflects the different occupation behavior of the quons. The
two body pairing interaction is  modified by the presence of the
factor $ (1+q^{4})/2$. In the fermionic case,  the $H^{\prime}$ term
is proportional to the difference between the total number of
pairs  and the total pair degeneracy,
\begin{equation}
H'_{fermions}= -G[\Omega - {(\hat N_1 + \hat N_2)}].
\label{hlinhaf}
\end{equation}
while in the quonic case it displays clearly the modification on
the occupation behavior of the quon particles giving a different
weight for  the number of pairs. When $q=-1$ we
obtain the right fermionic limit for the pairing hamiltonian.

\subsection{Two level pairing model solution}

Before defining the basis states needed to solve
the above hamiltonian, it is worth mentioning that the operator
introduced in equation (\ref{Azao}) is a tensor operator of order
zero once, as shown in \cite{su2}, quons obey the usual angular
momentum coupling rules. So we can talk about a state formed by N
zero angular momentum  quon pairs (or $s$-quon pairs),  which
defines the ground state. In other words, a system of N $s$-quon
pairs should be of the form ${\cal A}(\hat A^{\dag})^N|0>$. In
fact, using the properties presented in the last section, one can
show that the normalized antisymmetric $N$ quon pair state is given
by:
\begin{equation}
|N;A>_{\cal N}=\frac{\Omega^{N/2}}{N!}{\cal A}(\hat
A^{\dag})^N|0>.
\end{equation}

For the two level case, except for a normalization factor, our
basis states are
\begin{equation}
|N_i~N_k;A>\simeq {\cal A}(\hat A^{\dag}_i)^{N_i}(\hat
A^{\dag}_k)^{N_k}|0>,
\end{equation}
where $i$ and $k$ can be either 1 or 2 and $N=N_i+N_k$. The
calculation of the matrix elements can be done using our results
shown in section 2, which is a tedious but straightforward procedure.
For the number operator we get

\begin{equation}
<A;N'_i~N'_j|\hat N_i|N_i~N_j;A> = \frac{\{2N\}}{2N} N_i
\delta_{N'_{i} N_{i}} \delta_{N'_{j} N_{j}}.
\end{equation}
When $q\neq-1$ this result is different from $N_{i}$, the number
of pairs in the level $ i$. Thus the usual  pair number  operator
$\hat{N}$ when applied to the $N$ quon pair state gives us a number
that reflects the different  quonic occupation scheme.  We are
going to interpret this result as an effective number of pairs or
the number of quasi-pairs. Applying now the operator $\hat A_i$
onto $|N_i~N_k;A>$ yields
\begin{equation}
\hat A_i|N_i~N_k;A>_{\cal N}= \sqrt{
\frac{\{2N\}\{2N-1\}}{2N(2N-1)}}
\sqrt{\frac{N_i(\Omega-N_i+1)}{\Omega}} |N_i-1~N_k;A>_{\cal N}.
\end{equation}
From the above equations, a general expression for the pairing
hamiltonian matrix elements can be written:
$$
<A;N'_1~N'_2|H|N_1~N_2;A>=\left[ \epsilon (N_1-N_2)
\frac{\{2N\}}{2N} - \frac{G}{2} \frac{ \{2N\} \{2N-1\}}{2N
(2N-1)}(1+q^4) \times \right.
$$
$$ \left.
\left(N_1(\Omega-N_1)+N_2(\Omega-N_2) + \Omega \right) + <H'>
\right] \delta_{N_1N'_1} \delta_{N_2N'_2}
$$
$$
-\frac{G}{2} \frac{ \{2N\} \{2N-1\}}{2N (2N-1)}(1+q^4) \times
$$
$$
\left[
\sqrt{(N_1+1)(\Omega-N_1)(\Omega-N_2+1)N_2}\delta_{N_1+1N'_1}
\delta_{N_2-1N'_2} \right.
$$
\begin{equation}
\left. +
\sqrt{(N_2+1)(\Omega-N_2)(\Omega-N_1+1)N_1}\delta_{N_1-1N'_1}
\delta_{N_2+1N'_2} \right], \label{diag}
\end{equation}
where
\begin {equation}
<H'>=\frac{G}{8}(q-1)[4\Omega+2(q-q^2)N].
\end {equation}

Again, in the limit $q=-1$ the hamiltonian matrix reduces to the
usual expression. Also, in that limit the term $<H'>$ goes to zero
when $N=\Omega$, which is no longer true for other $q$ values.

In figure 1 we have plotted the ground state energy
obtained from the diagonalization of the pairing hamiltonian as a function
of the
particle angular momentum $j$, for some selected values of $q$. We
have chosen $\epsilon=1$ and $G=0.3$ for the calculations and
considered the case $N=\Omega$, which means that in the absence of the
interaction the ground state is formed by the lower level
completely full . One can immediately see that the system becomes
less bound when the deformation is turned on, which is
true for any value of the interaction strength $G$, as we have
checked. To better understand the effect of the deformation on the
various terms in the hamiltonian, we have plotted in figure 2 the
ground state energy when we deform only the kinetic term,
the kinetic plus the interaction term and finally disregarding the
term $<H'>$ in the hamiltonian matrix. We clearly see that
the main responsible for the deformation effects is the
interaction term.

A quantity of crucial importance in the pairing interaction theory is the so
called gap energy between the ground state and the two
quasiparticle excitation state \cite{RingSchuck}.
In the one-level pairing model, which is mathematically simpler than the
two-level version, one can already understand the implications of using
quons instead of fermions in the system under consideration.
For simplicity, in what follows, we investigate the effects of the quon
algebra on the gap energy in the one-level pairing model.

\subsection{One level pairing interaction and the energy gap}

The restriction of our previous results for N pairs in a single
$j$-shell can be done taking $\epsilon=0$ and disregarding all the
terms with the subscript $2$ in the hamiltonian. The ground state energy
in the one level pairing model reads:
$$
E_0=<N;A|H|N;A>=-\frac{G}{2}\frac{\{2N\}\{2N-1\}}{2N(2N-1)}N(\Omega-N+1)
(1+q^4)
$$
\begin{equation}
+\frac{G}{8}(q-1)[4\Omega+2(q-q^2)\frac{\{2N\}}{2N}N].
\label{onel0}
\end{equation}
Before proceeding with the calculations, it is worthwhile to note the
effect of the factor
$$F(N)=\frac{\{2N\}\{2N-1\}}{2N(2N-1)},$$
where $\{2N\}\{2N-1\}$ gives the effective number of interactions among pairs
and $2N(2N-1)$ gives the number of interactions among pairs. The factor $F(N)$
measures the deviations due to the fractional quon occupation scheme.
This quantity is always smaller
than one and it depends on $q$  and $N$, being equal to one when
$q=-1$. The presence of this term tends to decrease the pairing
interaction.

To obtain the gap energy, as mentioned in last subsection, we need to
define the rank-$J$ tensor in terms of the quon operators:
\begin{equation}
\hat B^{\dag}_{JM} =\frac{1}{\sqrt{2}} \sum_{m_1 m_2}
(jm_1jm_2|JM) a^{\dag}_{m_1} a^{\dag}_{m_2}.
\end{equation}
This operator creates a pair with angular momentum  $J$ and can be
used to define a one broken pair  excited state .
\begin{equation}
|N-1,1;A>_{\cal N}={\cal N_{J}}{\cal A}(\hat A^{\dag})^{N-1}\hat
B^{\dag}_{JM}|0>,
\end {equation}
${\cal N_{J}}$ being a normalization factor. The energy gap is
then given by
$$ \Delta =E_2-E_0=<N-1,1;A|H|N-1,1;A>-<N;A|H|N;A>=$$
\begin{equation}
\frac{G\Omega}{2}(1+q^4)\frac{\{2N\}\{2N-1\}}{2N(2N-1)},
\label{onel1}
\end{equation}
which can be compared with the usual value \cite{RingSchuck}. We can see
that the energy for breaking a pair is
now proportional to the factor ${(1+q^4)}/{2}$ which comes from
the symmetrization of the two-body interaction term of the
hamiltonian with the quonic commutation rules, and it is
also proportional to $F(N)$. As mentioned before this fraction is
always smaller than one and clearly stress the fact that the
energy necessary to break a pair of quons in a fully antisymmetric
state depends on the total number of particles in the system. In
other words, the occupation scheme of the quons introduces a
medium dependence on the gap. In order to better understand the effect of the
quon occupation scheme it is convenient to redefine the $q$
parameter as a function of $\Omega$, the degeneracy of the level,
where $q\sim -1$. We take

\begin{eqnarray}
 |q|=x^{1/\Omega}
\end{eqnarray}
so that eq. (\ref{anticaixote}) can be rewritten as
\begin{eqnarray}
 \{n\}=\frac{1-x^{n/\Omega}}{1-x^{1/\Omega}}
\end{eqnarray}
with $0\leq x \leq 1$. Using this definition we have
plotted the value of $\Delta /G\Omega$ as a function of the
deformation parameter $x$ in figure 3 for a fixed value of $\Omega
=40$  and different number of particles. Reminding that $\Delta
/G\Omega=1$  for the fermionic case ($x=1$) the departure from the
horizontal line exhibits the effect of the quonic deformation.
When the number of pairs approaches the degeneracy $\Omega$, the r\^ole
played by the deformation is more evident because the correlations due
to the occupation scheme becomes more important. If we choose another value for
$\Omega$ we have qualitatively the same trend shown in figure 3.

Finally, we may now establish a link between our results for the
energy gap behavior and the results for the ground state energy in
the two level model: once the deformation conspires against the
formation of Cooper pairs, the binding energy of the system tends
to become smaller than in the usual fermionic case.

\section{ Conclusions}

We have extended some of the results obtained in
\cite{sidric} to the permutational antisymmetric part of the whole
quonic space, which allows us to restrict the dynamics to a system
of fermion-like identical particles, in the same way we have
restricted it to boson-like particles using only the symmetric
subspace. A general expression for the annihilation of a particle
from a $N$ quon antisymmetric state was found, so that any
observable can be determined within the corresponding
subspace. An application of our results was made to study the
behavior of a system of quons interacting through a pairing force.
The main conclusion is that the Cooper pair formation, which is
characteristic of that type of interaction, can be largely
weakened for small deviations from a fermionic system. This
behavior can be attributed to the fact that the deformation
introduces an extra dependence in the two-body force which becomes
more and more important with the increase of the total number of
particles in the system and tends to effectively decrease the interaction.
This seems to be a property obeyed by
interacting quons in general and not only due to the pairing
interaction, but this is an investigation that remains to be confirmed.

A possible interesting problem is the consideration of a system formed by a
gas of particles with half-integer spin that are composed by many
fermions and that interact through the pairing force. In analogy
to what was discussed in \cite{Lipkin} for composed bosons, the
commutation relations obeyed by creation and annihilation
operators that define those particles are not the same as the ones obeyed by a
fermion (or a boson) and the departure from the
fermionic behavior can be described by the quon algebra.
This could be, for instance, the case in the recent
experiments with "fermion" traps \cite{Amusia} in a high density
regime. In that case the fermions are in fact complex atoms and
possible pairing type interactions between them \cite{Mottelson}
are supposed to be responsible for some of measurable behaviors of the
gas. These systems can be, in principle, modelled by extensions of
the present results.

\vspace{0.5cm}
\begin{center}
{\bf Acknowledgements}
\end{center}

This work was partially supported by  CNPq.

\begin{figure}
\begin{center}
\epsfig{file=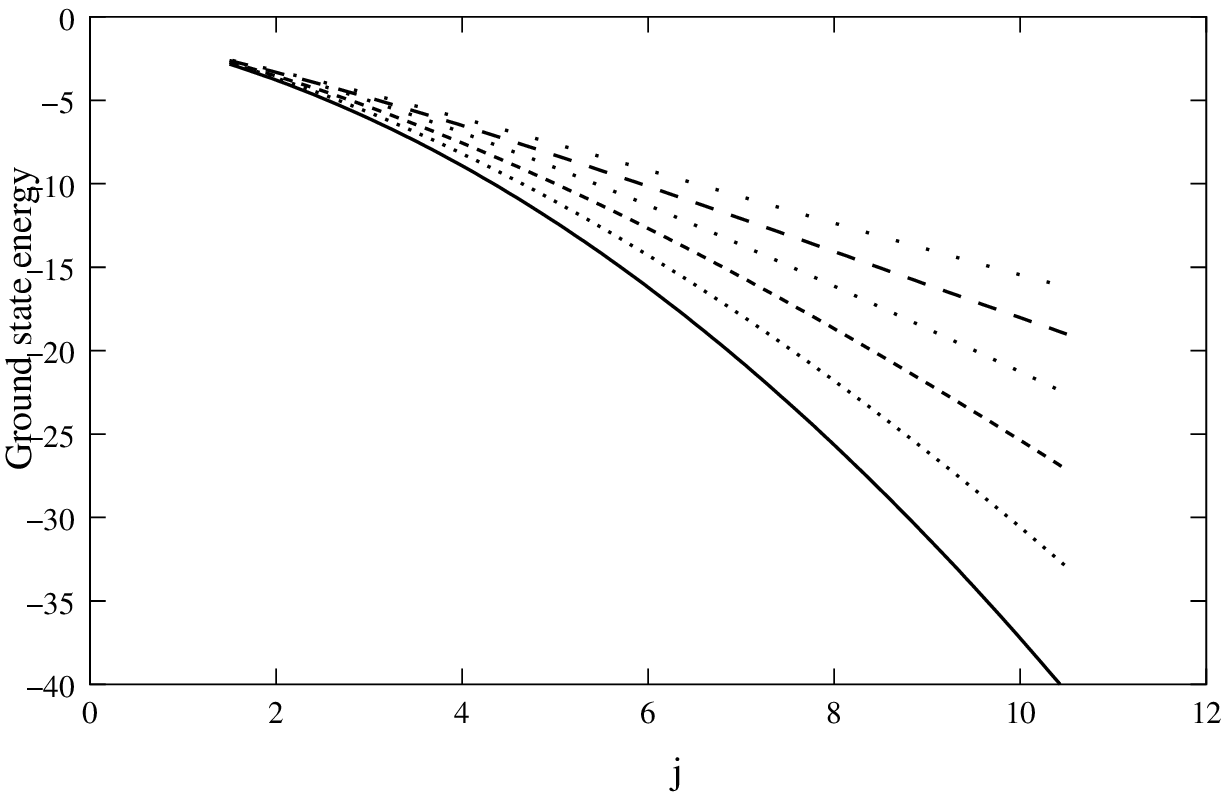} \caption{Ground state energy for $G=0.3$ and
$\epsilon=1$. The curves drawn from bottom to top were obtained
respectively with $q=-1$, $q=-0.99$, $q=-0.98$, $q=-0.97$,
$q=-0.96$ and $q=-0.95$. The energy is given in arbitrary units.}
\label{fig1}
\end{center}
\end{figure}

\begin{figure}
\begin{center}
\epsfig{file=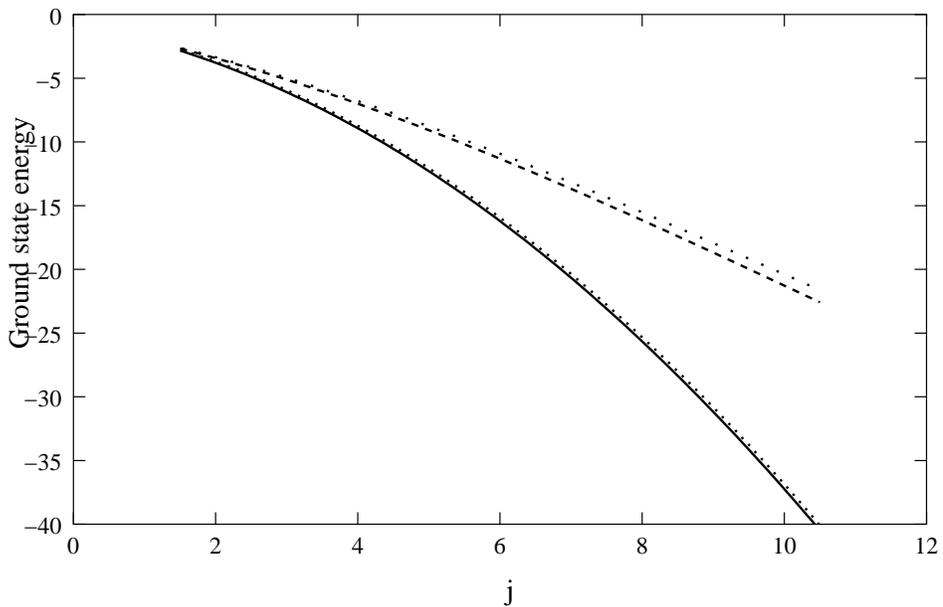} \caption{Ground state energy for $G=0.3$
and  $\epsilon=1$. The curves drawn from bottom to top were
obtained respectively with $q=-1$, $q=-0.97$ deforming only the
kinetic term, $q=-0.97$ deforming all terms of the hamiltonian and
$q=-0.97$ with $<H'>=0$. The energy is given in arbitrary units.}
\label{fig2}
\end{center}
\end{figure}

\begin{figure}
\begin{center}
\epsfig{file=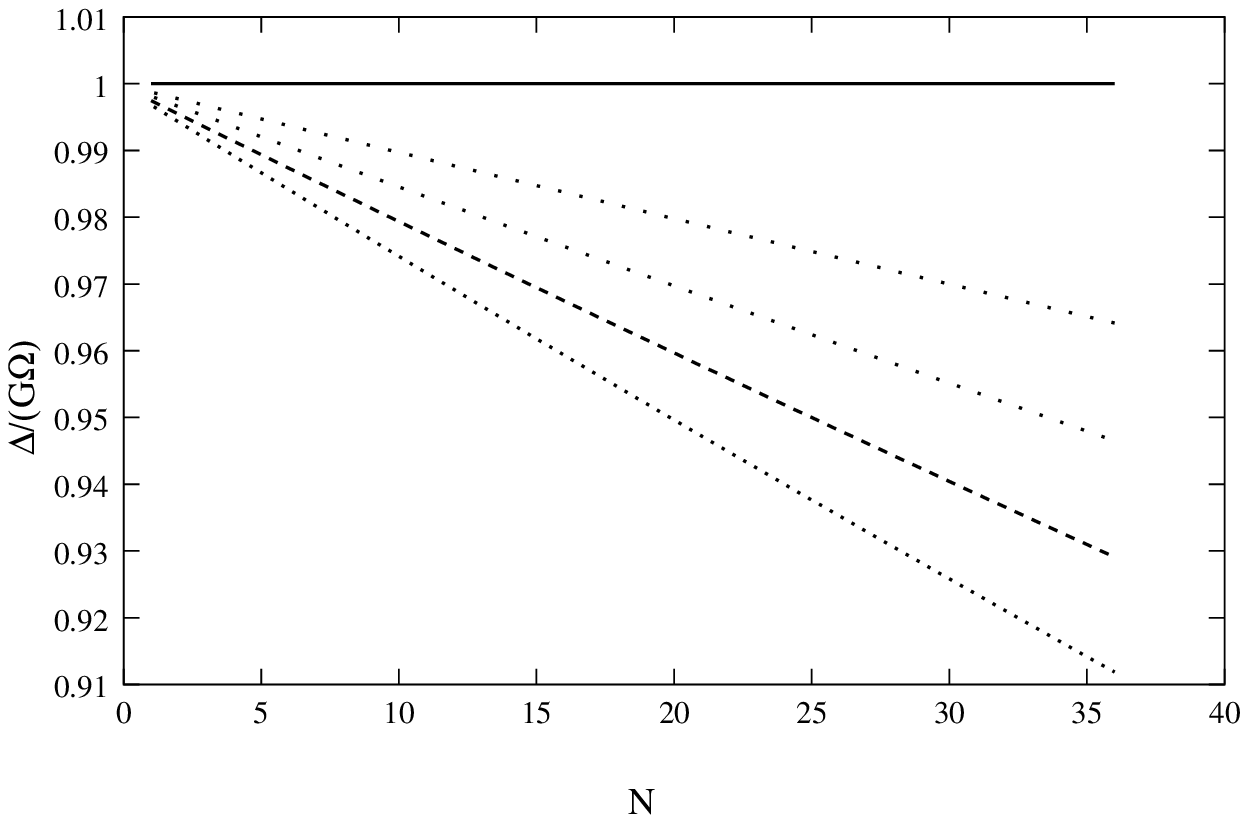} \caption{ Energy gap in units of $G\Omega$
as a function of the number of pairs $N$ with $\Omega=40$. The
curves drawn from bottom to top are for $x=0.96, 0.97, 0.98, 0.99$
and $1.0$ respectively.} \label{fig3}
\end{center}
\end{figure}


\begin{thebibliography}{99}

\bibitem{quons} O.W. Greenberg, Phys. Rev. {\bf D 43} (1991) 4111; M.
Chaichian, R. Gonzalez Felipe and C. Montonen, J. Phys. {\bf A 26} (1993)
4017.

\bibitem{sidric} S.S. Avancini, J.R. Marinelli and C.E. de O. Rodrigues,
Phys. Lett. {\bf A 297} (2002) 137.

\bibitem{keldysh} A. Griffin, D.W. Snoke and S. Stringari, Bose-Einstein
Condensation, Cambridge University Press, Cambridge:1995, L.V. Keldysh, 246-280; J.P. Wolfe, J.L. Lin and D.W. Snoke, 281-329; A. Mysyrowics, 330-354.

\bibitem{cambia}  M.C. Cambiaggio, G.G. Dussel and M. Saraceno, Nucl.
Phys. {\bf A 415} (1984) 70

\bibitem{Duzzioni} E.I. Duzzioni and J.R. Marinelli, Depto. de F\'{\i}sica,
Universidade Federal de Santa Catarina - private communication.

\bibitem{nossos} S.S. Avancini, F.F. de Souza Cruz, J.R. Marinelli and
D.P. Menezes, Phys. Rev. {\bf C 62} 024312; S.S. Avancini and D.P.
Menezes, J. Phys. {\bf A 26} (1993) 6261.

\bibitem{su2} S.S. Avancini, F.F. de Souza Cruz, J.R. Marinelli and D.P.
Menezes, Phys. Lett. {\bf A 267} (2000) 109.

\bibitem{RingSchuck} P. Ring and P. Schuck, The Nuclear Many-Body
Problem, Springer-Verlag (1980).

\bibitem{Lipkin} H. J. Lipkin, Quantum Mechanics, Elsevier (1973); S.S.
Avancini
and G. Krein, J. Phys. {\bf A 28} (1995) 685.

\bibitem{Amusia} M. Ya. Amusia, A.Z. Msezane and V.R. Shaginyan,
Phys. Lett. {\bf A 293} (2002) 205.

\bibitem{Mottelson} H. Heiselberg and B. Mottelson,
cond-mat/0112248 v1, 2001.

\end{thebibliography}
\end{document}